\documentclass[9pt,twocolumn,twoside]{opticajnl}
\journal{opticajournal} 

\setboolean{shortarticle}{true}

\usepackage{lineno}
\usepackage{times}
\usepackage{braket}

\title{High-efficiency, cryogenic-compatible grating couplers on an AlN-on-sapphire platform through bottom-side coupling}

\author[1]{Yiyu Zhou}
\author[1]{Mohan Shen}
\author[1]{Chunzhen Li}
\author[1]{Likai Yang}
\author[1]{Jiacheng Xie}
\author[1,*]{Hong X. Tang}

\affil[1]{Department of Electrical Engineering, Yale University, New Haven, Connecticut 06520, USA}
\affil[*]{hong.tang@yale.edu}

\begin{abstract}
Sapphire is a commonly used substrate for wide-bandgap III-nitride photonic materials. However, its relatively high refractive index results in low transmission efficiency in grating couplers. Here, we propose and demonstrate that the transmission efficiency can be significantly enhanced by bottom-side coupling. A metal reflector is deposited on the top side of the chip, and the fiber array is glued to the bottom side of the substrate. We experimentally achieve a transmission efficiency as high as 42\% per coupler on an aluminum nitride (AlN) on sapphire platform at the telecom wavelength. In addition, the grating couplers show a robust performance at a cryogenic temperature as low as 3~K for both transverse-electric (TE) and transverse-magnetic (TM) modes. Our results can be useful to a wide range of sapphire-based applications that require low coupling loss and cryogenic operation.
\end{abstract}

\setboolean{displaycopyright}{false} 

\begin{document}

\maketitle

Grating couplers play a crucial role for integrated silicon photonics as an efficient and compact interface for fiber-to-chip coupling. Thanks to the large refractive index contrast between silicon and silicon dioxide, the transmission efficiency on a silicon-on-insulator wafer has approached 50\% per coupler \cite{zemtsov2022broadband}, and it can be further improved to >80\% per coupler by using a bottom metal reflector \cite{zaoui2014bridging, ding2014fully, hoppe2019ultra}. On the other hand, wide-bandgap III-nitride semiconductors, such as aluminum nitride (AlN) and gallium nitride (GaN), are emerging as integrated photonics materials due to the presence of second-order optical nonlinearity. In particular, AlN has been extensively investigated as a promising platform for a variety of nonlinear optical applications \cite{liu2023aluminum, li2021aluminium}, including electro-optic modulation \cite{xiong2012low, zhu2016aluminum}, microwave-to-optical transduction \cite{mirhosseini2020superconducting, fan2018superconducting, lecocq2021control, fu2021cavity, holzgrafe2020cavity, mckenna2020cryogenic, rueda2016efficient}, quantum optics \cite{guo2017parametric}, second-harmonic generation \cite{bruch201817}, soliton and micro-comb generation \cite{yao2021pure, liu2021raman, bruch2021pockels, liu2020photolithography, afridi2022breather}, Raman lasers \cite{liu2022fundamental, liu2017integrated}, and optical frequency shifting \cite{fan2016integrated}. However, for the consideration of lattice matching, high-quality single-crystalline AlN is typically deposited on the sapphire substrate. The low refractive index contrast between AlN ($n \approx 2.1$) and sapphire ($n \approx 1.7$) has thus impeded the development of efficient AlN-on-sapphire grating couplers, and the transmission efficiency is lower than 25\% per coupler even in simulation \cite{lu2018aluminum}. Attempts have been made to enhance the transmission efficiency by applying a silicon layer on AlN, and the simulated efficiency can achieve 60\% per coupler, while the experimentally measured efficiency is reported to be 28\% per coupler \cite{Shreelakshmi22high}. A commonly used method to enhance the transmission efficiency in silicon photonics is to use a bottom metal reflector \cite{zaoui2014bridging, ding2014fully, hoppe2019ultra} by etching the entire silicon substrate via deep reactive ion etching. However, this method is not applicable to sapphire because the sapphire substrate is both chemically and physically stable and thus does not allow efficient deep etching. Metal grating couplers \cite{smith2022toward, ruan2020metal} present another approach to high coupling efficiency. However, metal grating couplers are not compatible with the high-temperature annealing process, which is often used to improve the optical quality factors of AlN resonators \cite{fan2018superconducting}. Edge coupling is another efficient method for AlN photonics, and a coupling loss of 2.8 dB per facet has been demonstrated \cite{liu2017aluminum}. However, edge coupling requires a large footprint, precise position alignment, and smoothly cleaved edges that are challenging for the sapphire substrate. Therefore, high-efficiency grating couplers remain highly desirable for AlN-on-sapphire photonics.

In this work, we propose and demonstrate a bottom-side coupling technique that can significantly enhance the transmission efficiency. We deposit a metal reflector on the top side of the grating couplers such that most light is scattered downward by the grating. An apodized grating design is adopted to focus the scattered light to the fiber array that is glued underneath the 430-\textmu m-thick sapphire substrate. The transmission efficiency is measured to be as high as 42\% per coupler. The grating coupler also presents similar transmission for both TE and TM modes simultaneously. We further test the robustness of the grating coupler at cryogenic temperatures by cooling down to 3~K, and the coupling efficiency remains nearly unchanged. The compatibility with the low temperature is useful to cryogenic applications such as microwave-to-optical transducers \cite{mirhosseini2020superconducting, fan2018superconducting, lecocq2021control, fu2021cavity, holzgrafe2020cavity, mckenna2020cryogenic, rueda2016efficient} and superconducting electro-optic modulators \cite{youssefi2021cryogenic, shen2024photonic}, where the photonic chip needs to be placed in a cryogenic environment to enable the operation of superconducting circuits.

\begin{figure}[t]
\centering
\includegraphics[width=\linewidth]{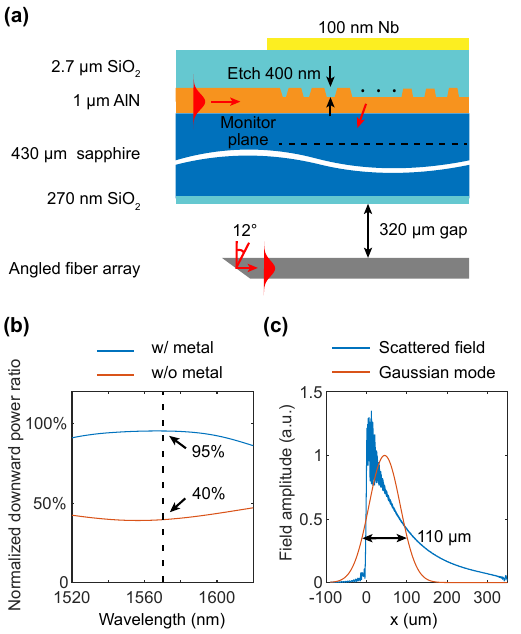}
\caption{(a) The schematic of the AlN-on-sapphire grating coupler. The monitor plane is placed 3 \textmu m under the AlN film in the simulation program. (b) The simulated normalized downward power ratio as a function of wavelength. (c) The simulated field amplitude distribution of the scattered field and the Gaussian fiber mode at the monitor plane. (d) The normalized transmission per coupler as a function of position misalignment.}
\label{fig:schematic}
\end{figure}

\textbf{Grating coupler design.} The design of the grating coupler is shown in Fig.~\ref{fig:schematic}(a). The AlN film thickness is 1~\textmu m, and the grating etching depth is 400~nm. The sapphire substrate is double side polished and has a thickness of 430~\textmu m. A 270-nm-thick SiO$_2$ layer is deposited at the bottom side as an anti-reflection coating. We use 2.7-\textmu m-thick SiO$_2$ as the cladding, and 100~nm niobium (Nb) is deposited on the grating as a metal reflector. An angled fiber array is glued at the bottom side of the wafer. The light in the fiber is reflected towards the grating at the fiber tip due to the total internal reflection at the fiber-air interface. To illustrate the effect of the metal reflector, we launch the TM0 mode in the on-chip waveguide and numerically calculate the downward scattering power ratio with and without the metal reflector using a commercial software (Ansys Lumerical FDTD) in a two-dimensional (2D) configuration. The normalized downward scattering power ratio is presented in Fig.~\ref{fig:schematic}(b). In the absence of a metal reflector, only $\sim$40\% of the light is directed to the fiber. By contrast, when a metal reflector is present, the downward scattering power ratio increases to as high as 95\%, which validates the use of metal reflector. We note that we design the grating coupler for TM0 mode because we experimentally observed that the TM0 mode has higher quality factor compared to the TE0 mode.

We next investigate the design of grating structure. Each grating tooth can be viewed as a scatterer that directs a small amount of light to the fiber \cite{zhao2020design}. The scattering strength depends on several parameters such as the index contrast, the duty cycle, and the etching depth. In general, a high scattering strength is desirable because it allows a smaller scattered field size to match the fiber mode diameter. We simulate different etching depths for AlN on sapphire and eventually choose an etching depth of 400~nm, because a larger etching depth does not further enhance the scattering strength significantly. It is worth noting that the scattering strength of the grating can be appreciably tuned by adjusting the grating duty cycle in silicon photonics, thanks to the high refractive index contrast. Hence, by spatially varying the duty cycle, an apodized grating coupler can be designed to generate a Gaussian-like scattered field profile to match the fiber mode \cite{zhao2020design}. However, for AlN-on-sapphire platforms, we find that the scattering strength does not show significant dependence on the grating duty cycle as a consequence of the low index contrast, and thus the scattered field always shows a negative exponential distribution. The maximum coupling efficiency is therefore upper bonded to 80\%, which is determined by the overlap between a Gaussian distribution and an exponential distribution \cite{zhao2020design}. To match the fiber Gaussian mode size to the scattered field size, we place the fiber array 320~\textmu m underneath the sapphire bottom side. The fiber is polished at an angle of $41^\circ$, and thus the reflected fiber mode propagates at an angle of $12^\circ$ with respect to the $z$ axis in air. The diffraction angle of the grating is hence chosen to be $12^\circ$ to match the angled fiber as shown in Fig.~\ref{fig:schematic}(a).

\begin{figure}[t]
\centering
\includegraphics[width=\linewidth]{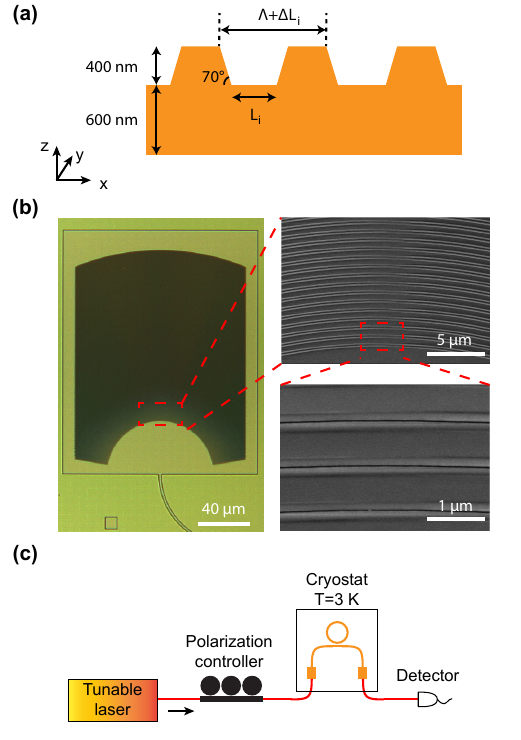}
\caption{(a) Design parameter of the grating coupler. (b) The optical micrograph and the scanning electron microscope (SEM) image of a grating coupler. The images are taken before the PECVD SiO$_2$ cladding is deposited. (c) The grating coupler measurement setup.}
\label{fig:micrograph}
\end{figure}

\begin{figure}[t]
\centering
\includegraphics[width=\linewidth]{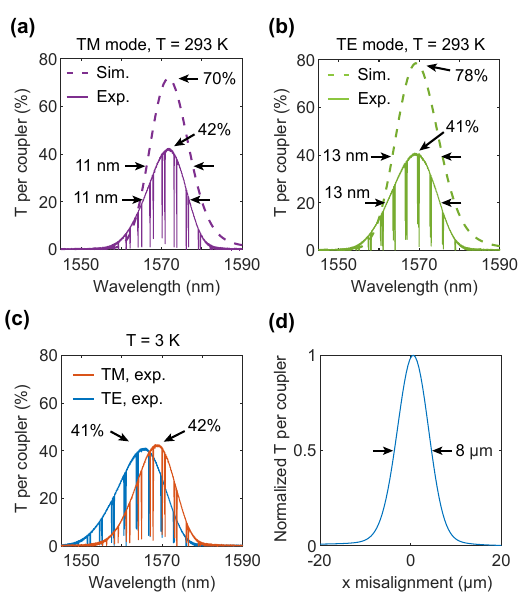}
\caption{The simulated and experimentally measured transmission per coupler for (a) TM mode and (b) TE mode at room temperature T=293~K. (c) The experimentally measured transmission per coupler for TE and TM modes at cryogenic temperature T=3~K. (d) The normalized transmission per coupler for position misalignment in the $x$ dimension.}
\label{fig:transmission}
\end{figure}

In the numerical simulation, we design the grating coupler for TM polarization and we launch the TM0 mode in the waveguide as the input mode. A monitor plane is placed 3~\textmu m under the grating coupler. The scattered field profile is shown in Fig.~\ref{fig:schematic}(c). It can be seen that the scattered field size is significantly larger than the fiber mode field diameter of 10.6~\textmu m. Therefore, we place the fiber 320~\textmu m underneath the sapphire substrate. After propagation in air gap and sapphire substrate, the fiber mode field diameter at the monitor plane becomes as large as 110~\textmu m as a consequence of divergence and thus matches the scattered field size. The fill factor and grating period are tuned to determine the diffraction angle $\theta$, and the relation can be written as \cite{lomonte2021efficient}
\begin{equation}
\begin{split}
    \Lambda &= \frac{\lambda}{n_{\text{eff}}+ \sin \theta}, \\
    n_{\text{eff}}&=n_{\text{wg}}\cdot F + n_{\text{e}}\cdot (1-F),
\end{split}
\end{equation}
where $\lambda$ is the vacuum wavelength, $\Lambda$ is the grating period, $n_{\text{eff}}$ is the effective index of a period, $F$ is the fill factor, $n_{\text{wg}}=1.99$ is the eigenmode effective index in the unetched waveguide, and $n_{\text{e}}=1.90$ is the eigenmode effective index of the etched waveguide. Due to the birefringence of single-crystalline AlN, the refractive index of AlN used in the simulation is $n_{\text{TM}}=2.050$ and $n_{\text{TE}}=2.016$ for TM and TE polarization, respectively. In our design, we use $\theta=12^{\circ}$ to match the beam deflection angle of the angled fiber as shown in Fig.~\ref{fig:schematic}(a), and the fill factor starts from 0.85 at $x=0$ and linearly decreases to 0.70 at $x=30$~\textmu m, then remains at 0.70 for $x>30$~\textmu m. Here $x=0$ is defined as the position of the first grating unit cell. To focus the scattered field to the fiber, we use an apodized design \cite{zhao2020design} based on spatially varying grating period. For a focusing Gaussian mode centered at $x_c$, its phase distribution can be written as $\phi(x) = -2\pi(x-x_c)^2/(2R_z\lambda)$, where $x_c$ is the center of the fiber Gaussian mode at the grating plane, $R_z$ is the radius of curvature of the wavefront at the grating plane and is determined by $R_z=l+(z_R^2/l)$, $z_R=\pi \omega_0^2/\lambda$ is the fiber mode Rayleigh range, $\omega_0= 5.2$ \textmu m is the fiber beam waist radius, $l=(t_{\text{sa}}/n_{\text{sa}})+t_{\text{air}}$ is the equivalent path length in air between the fiber and the grating coupler, $n_{\text{sa}}=1.74$ is the sapphire substrate refractive index, $t_{\text{sa}}=430$ \textmu m is the sapphire substrate thickness, and $t_{\text{air}}=320$ \textmu m is the air gap thickness. The value of $x_c$ needs to be numerically optimized, and we use $x_c=43$ \textmu m in our simulation. To imprint this phase distribution to the scattered field, we add a small length change $\Delta L$ to each grating period, which follows \cite{zhao2020design}
\begin{equation}
\Delta L_i= \frac{\lambda}{2\pi ( n_{\text{wg}} + \sin \theta )  } (\phi_i - \phi_{i-1}),     
\end{equation}
where $\Delta L_i$ is the length change in the $i$-th unit cell, $\phi_i=\phi(x_i)$ is the phase at the position of the $i$-th unit cell $x_i$. Therefore, the total period of the $i$-th unit cell is $\Lambda + \Delta L_i$, and the length of the etched part is $L_i=\Lambda (1-F)$, as depicted in Fig.~\ref{fig:micrograph}(a).

\textbf{Fabrication and measurement.} 
The 1-\textmu m-thick AlN film is grown on double-side polished sapphire substrate with a thickness of 430~\textmu m using metal-organic chemical vapor deposition (MOCVD). To fabricate the grating, we deposit 175~nm SiO$_2$ on AlN as a hard mask by plasma-enhanced chemical vapor deposition (PECVD). Electron beam resist (CSAR 62) is then spin coated onto the SiO$_2$ layer, and 10~nm gold is sputtered subsequently as a charge dissipation layer \cite{wang2024heterogeneous}. The resist is exposed by a 100~kV electron beam pattern generator (EBPG 5200, Raith). We remove the gold layer by dipping in gold etchant 
for 2 minutes and then develop the resist by dipping in xylene for 45 seconds. We etch the SiO$_2$ hard mask by CHF$_3$/O$_2$, and then etch the AlN layer by Cl$_2$/BCl$_3$/Ar in an Oxford 100 etcher \cite{liu2015smooth}. The remaining SiO$_2$ hard mask is removed by buffered oxide etch. The waveguides and micro-ring resonators are fabricated in a separate step with an etching depth of 600~nm. We note that the micro-ring resonators are used to inspect the waveguide propagation loss and polarization state on chip. The resonators have no effect on the performance of grating couplers and thus we do not further discuss their designs. 2.7~\textmu m SiO$_2$ is deposited on the grating by PECVD as the cladding. 100~nm Nb is deposited on the grating areas by electron beam evaporation, and the pattern of Nb is defined by a photolithography lift-off process. A layer of 270~nm PECVD SiO$_2$ is deposited at the bottom side of the sapphire substrate as an anti-reflection coating for telecom wavelengths. We note that the Nb reflector can be readily replaced by other commonly used metals such as gold and aluminum. The micrographs of the fabricated grating couplers are presented in Fig.~\ref{fig:micrograph}(b). The fiber array is attached to the bottom side following a recipe similar to \cite{shen2024photonic}. The device chip is first glued on a copper plate. A hole is drilled at the edge of the copper plate to make the chip bottom side accessible. The fiber array is aligned to the grating couplers and then glued to the chip bottom side using an ultraviolet-curable epoxy.

The measurement setup to characterize the grating couplers is shown in Fig.~\ref{fig:micrograph}(c). We use a wavelength-tunable laser (TSL-710, Santec) as the light source to characterize the spectral response of the grating couplers. A three-paddle fiber polarization controller (FPC560, Thorlabs) is used to tune the polarization state of light, and a telecom fiber-coupled detector (2053-FC, New Focus) is used to measure the optical power. For room-temperature measurement, we first short connect the fibers to bypass the chip and measure the optical power $P_0$. We then connect fibers to the chip and measure its spectral response $P_{rt}(\lambda)$ by sweeping the laser wavelength $\lambda$. We then place the chip in a cryostat and measure the cryogenic response  $P_{c}(\lambda)$. Each device on the chip has two grating couplers, one as input port and the other as output port. We assume that two grating couplers are identical, and thus the transmission per coupler can be computed as $\sqrt{P_{rt}(\lambda)/P_0}$ and $\sqrt{P_{c}(\lambda)/P_0}$, respectively. The simulated as well as the experimentally measured transmission results for at room temperature are presented in Fig.~\ref{fig:transmission}(a) for TM mode and Fig.~\ref{fig:transmission}(b) for TE mode. Resonance dips are visible in the measured spectrum due to the presence of micro-ring resonators. These resonances are useful for distinguishing the polarization states and have no effect on the coupling efficiency measurement. The simulation shows a maximum transmission of 70\% (78\%) per coupler, while the measured transmission is 42\% (41\%) per coupler for TM (TE) mode. We attribute the slightly higher simulated efficiency of TE mode to its stronger grating scattering strength observed in the simulation, and the different peak wavelength is attributed to the different effective mode index and material birefringence. Regarding the discrepancy in efficiency between simulation and experiment, there are several possible reasons. First, the simulation is performed in a simplified 2D configuration along the radial dimension for time-efficient computations. We expect a three-dimensional (3D) simulation to produce a lower coupling efficiency due to the mode mismatch along the angular dimension. Second, we control the AlN etching depth by measuring the AlN thickness using an ellipsometer. However, an ellipsometer can only measure the thickness of a uniform film. The plasma etching process generally shows an aspect ratio dependence, and we expect the etching depth at the grating area to be smaller than the value measured by an ellipsometer \cite{yeom2005maximum}. In addition, the PECVD cladding is known to produce air voids at grating gaps \cite{sun2019ultrahigh}, which can change the grating scattering strength unpredictably. The etching depth and air voids can potentially be characterized by inspecting the cleaved side of a chip using SEM, which we leave for future study. We also characterize the performance of six different grating couplers, and the average efficiency is 41.4\% with a standard deviation of 1.6\%, which validates the reproducibility of the structure. The tolerance to the fiber position misalignment is also estimated numerically. Based on the simulated scattered field distribution $E_s(x)$ and the fiber Gaussian mode $E_f(x)$ at the monitor plane (see Fig.~\ref{fig:schematic}(c)), the transmission under misalignment can be computed as $ \left|\int E_s^*(x) E_f(x-\delta x)dx\right|^2$, where $\delta x$ is the position misalignment in the $x$ dimension. The results show a 3~dB tolerance of 8~\textmu m (see Fig.~\ref{fig:transmission}(d)).

To test the robustness of grating couplers at low temperatures, We put the chip in a cryostat which allows to cool down to T=3~K within 8 hours. The peak transmission is nearly unchanged, and the peak wavelength shows a slight shift of 3~nm as demonstrated in Fig.~\ref{fig:transmission}(b). We attribute the peak wavelength shift to the thermal contraction of the sapphire substrate during the cool down process. The 3~dB bandwidth of the coupler is measured to be 11~nm for TM mode and 13~nm for TE mode, which agrees well with the simulation result. The limited bandwidth stems from the large distance between the fiber and the grating coupler \cite{lomonte2021efficient}, which fundamentally is a result of the low index contrast between AlN and sapphire. We note that this bandwidth is sufficient for microwave-to-optical transducers, and it could potentially be improved by using silicon overlay layer \cite{Shreelakshmi22high} to enhance the index contrast. Although the grating coupler is designed for TM polarization, we notice that the a similar peak transmission of 41\% can be achieved for TE polarization at T=3~K, which can be useful to applications that require both polarizations.

\textbf{Conclusions.} In summary, we experimentally demonstrate high-efficiency grating couplers on an AlN-on-sapphire platform via bottom-side coupling. The low index contrast between AlN and sapphire necessitates the large distance between the fiber and the grating for mode size matching. To enhance the scattering directionality, a metal layer is deposited on the top side of the grating coupler as a reflector, and the fiber array is hence attached to the bottom side of the double-side-polished substrate. An apodized design with spatially varying grating periods is adopted to focus the scattered field to the fiber. We experimentally achieved a peak transmission as high as 42\% per coupler for TM polarization, both at room temperature T=293~K and cryogenic temperature T=3~K, which allows for compatibility with superconducting quantum circuits \cite{mirhosseini2020superconducting, fan2018superconducting, lecocq2021control, fu2021cavity, holzgrafe2020cavity, mckenna2020cryogenic, rueda2016efficient}. In addition, the grating coupler simultaneously support the transmission of both TE and TM polarizations with similar peak transmission and bandwidth. We believe that our design is not only useful to AlN-on-sapphire platform, but also can benefit other low-index-contrast sapphire-based platform such as silicon nitride on sapphire \cite{martinussen2024thick} and lithium niobate on sapphire \cite{mckenna2020cryogenic, mishra2021mid} substrates.

\begin{backmatter}
\bmsection{Funding} Co-design Center for Quantum Advantage (DE-SC0012704); Army Research Office (W911NF2410029).

\bmsection{Acknowledgment} The authors thank Dr. Yong Sun, Dr. Michael Rooks, Dr. Lauren McCabe, Dr. Yeongjae Shin, and Kelly Woods for assistance in device fabrication. The authors acknowledge the support from Yale Quantum Institute, Yale University Cleanroom, and Yale Institute for Nanoscience and Quantum Engineering.

\bmsection{Disclosures} The authors declare no conflicts of interest.

\bmsection{Data availability} Data underlying the results presented in this paper are not publicly available at this time but may be obtained from the authors upon reasonable request.
\end{backmatter}

\bibliography{sample}

\begin{thebibliography}{10}
\newcommand{\enquote}[1]{``#1''}

\bibitem{zemtsov2022broadband}
D.~S. Zemtsov, D.~M. Zhigunov, S.~S. Kosolobov, \emph{et~al.}, \enquote{Broadband silicon grating couplers with high efficiency and a robust design,} {\protect\JournalTitle{Opt. Lett.}} \textbf{47}, 3339--3342 (2022).

\bibitem{zaoui2014bridging}
W.~S. Zaoui, A.~Kunze, W.~Vogel, \emph{et~al.}, \enquote{Bridging the gap between optical fibers and silicon photonic integrated circuits,} {\protect\JournalTitle{Opt. Express}} \textbf{22}, 1277--1286 (2014).

\bibitem{ding2014fully}
Y.~Ding, C.~Peucheret, H.~Ou, and K.~Yvind, \enquote{Fully etched apodized grating coupler on the {SOI} platform with -0.58 db coupling efficiency,} {\protect\JournalTitle{Opt. Lett.}} \textbf{39}, 5348--5350 (2014).

\bibitem{hoppe2019ultra}
N.~Hoppe, W.~S. Zaoui, L.~Rathgeber, \emph{et~al.}, \enquote{Ultra-efficient silicon-on-insulator grating couplers with backside metal mirrors,} {\protect\JournalTitle{IEEE J. Sel. Top. Quantum Electron.}} \textbf{26}, 1--6 (2019).

\bibitem{liu2023aluminum}
X.~Liu, A.~W. Bruch, and H.~X. Tang, \enquote{Aluminum nitride photonic integrated circuits: from piezo-optomechanics to nonlinear optics,} {\protect\JournalTitle{Adv. Opt. Photon.}} \textbf{15}, 236--317 (2023).

\bibitem{li2021aluminium}
N.~Li, C.~P. Ho, S.~Zhu, \emph{et~al.}, \enquote{Aluminium nitride integrated photonics: a review,} {\protect\JournalTitle{Nanophotonics}} \textbf{10}, 2347--2387 (2021).

\bibitem{xiong2012low}
C.~Xiong, W.~H. Pernice, and H.~X. Tang, \enquote{Low-loss, silicon integrated, aluminum nitride photonic circuits and their use for electro-optic signal processing,} {\protect\JournalTitle{Nano Lett.}} \textbf{12}, 3562--3568 (2012).

\bibitem{zhu2016aluminum}
S.~Zhu and G.-Q. Lo, \enquote{Aluminum nitride electro-optic phase shifter for backend integration on silicon,} {\protect\JournalTitle{Opt. Express}} \textbf{24}, 12501--12506 (2016).

\bibitem{mirhosseini2020superconducting}
M.~Mirhosseini, A.~Sipahigil, M.~Kalaee, and O.~Painter, \enquote{Superconducting qubit to optical photon transduction,} {\protect\JournalTitle{Nature}} \textbf{588}, 599--603 (2020).

\bibitem{fan2018superconducting}
L.~Fan, C.-L. Zou, R.~Cheng, \emph{et~al.}, \enquote{Superconducting cavity electro-optics: a platform for coherent photon conversion between superconducting and photonic circuits,} {\protect\JournalTitle{Sci. Adv.}} \textbf{4}, eaar4994 (2018).

\bibitem{lecocq2021control}
F.~Lecocq, F.~Quinlan, K.~Cicak, \emph{et~al.}, \enquote{Control and readout of a superconducting qubit using a photonic link,} {\protect\JournalTitle{Nature}} \textbf{591}, 575--579 (2021).

\bibitem{fu2021cavity}
W.~Fu, M.~Xu, X.~Liu, \emph{et~al.}, \enquote{Cavity electro-optic circuit for microwave-to-optical conversion in the quantum ground state,} {\protect\JournalTitle{Phys. Rev. A}} \textbf{103}, 053504 (2021).

\bibitem{holzgrafe2020cavity}
J.~Holzgrafe, N.~Sinclair, D.~Zhu, \emph{et~al.}, \enquote{Cavity electro-optics in thin-film lithium niobate for efficient microwave-to-optical transduction,} {\protect\JournalTitle{Optica}} \textbf{7}, 1714--1720 (2020).

\bibitem{mckenna2020cryogenic}
T.~P. McKenna, J.~D. Witmer, R.~N. Patel, \emph{et~al.}, \enquote{Cryogenic microwave-to-optical conversion using a triply resonant lithium-niobate-on-sapphire transducer,} {\protect\JournalTitle{Optica}} \textbf{7}, 1737--1745 (2020).

\bibitem{rueda2016efficient}
A.~Rueda, F.~Sedlmeir, M.~C. Collodo, \emph{et~al.}, \enquote{Efficient microwave to optical photon conversion: an electro-optical realization,} {\protect\JournalTitle{Optica}} \textbf{3}, 597--604 (2016).

\bibitem{guo2017parametric}
X.~Guo, C.-l. Zou, C.~Schuck, \emph{et~al.}, \enquote{Parametric down-conversion photon-pair source on a nanophotonic chip,} {\protect\JournalTitle{Light Sci. Appl.}} \textbf{6}, e16249--e16249 (2017).

\bibitem{bruch201817}
A.~W. Bruch, X.~Liu, X.~Guo, \emph{et~al.}, \enquote{17 000\%/{W} second-harmonic conversion efficiency in single-crystalline aluminum nitride microresonators,} {\protect\JournalTitle{Appl. Phys. Lett.}} \textbf{113} (2018).

\bibitem{yao2021pure}
S.~Yao, K.~Liu, and C.~Yang, \enquote{Pure quartic solitons in dispersion-engineered aluminum nitride micro-cavities,} {\protect\JournalTitle{Opt. Express}} \textbf{29}, 8312--8322 (2021).

\bibitem{liu2021raman}
K.~Liu, S.~Yao, and C.~Yang, \enquote{Raman pure quartic solitons in {Kerr} microresonators,} {\protect\JournalTitle{Opt. Lett.}} \textbf{46}, 993--996 (2021).

\bibitem{bruch2021pockels}
A.~W. Bruch, X.~Liu, Z.~Gong, \emph{et~al.}, \enquote{Pockels soliton microcomb,} {\protect\JournalTitle{Nat. Photon.}} \textbf{15}, 21--27 (2021).

\bibitem{liu2020photolithography}
J.~Liu, H.~Weng, A.~A. Afridi, \emph{et~al.}, \enquote{Photolithography allows high-{Q} {AlN} microresonators for near octave-spanning frequency comb and harmonic generation,} {\protect\JournalTitle{Opt. Express}} \textbf{28}, 19270--19280 (2020).

\bibitem{afridi2022breather}
A.~A. Afridi, H.~Weng, J.~Li, \emph{et~al.}, \enquote{Breather solitons in {AlN} microresonators,} {\protect\JournalTitle{Opt. Continuum}} \textbf{1}, 42--50 (2022).

\bibitem{liu2022fundamental}
K.~Liu, S.~Yao, Y.~Ding, \emph{et~al.}, \enquote{Fundamental linewidth of an aln microcavity raman laser,} {\protect\JournalTitle{Opt. Lett.}} \textbf{47}, 4295--4298 (2022).

\bibitem{liu2017integrated}
X.~Liu, C.~Sun, B.~Xiong, \emph{et~al.}, \enquote{Integrated continuous-wave aluminum nitride raman laser,} {\protect\JournalTitle{Optica}} \textbf{4}, 893--896 (2017).

\bibitem{fan2016integrated}
L.~Fan, C.-L. Zou, M.~Poot, \emph{et~al.}, \enquote{Integrated optomechanical single-photon frequency shifter,} {\protect\JournalTitle{Nat. Photon.}} \textbf{10}, 766--770 (2016).

\bibitem{lu2018aluminum}
T.-J. Lu, M.~Fanto, H.~Choi, \emph{et~al.}, \enquote{Aluminum nitride integrated photonics platform for the ultraviolet to visible spectrum,} {\protect\JournalTitle{Opt. Express}} \textbf{26}, 11147--11160 (2018).

\bibitem{Shreelakshmi22high}
S.~KP, S.~Raghavan, and S.~K. Selvaraja, \enquote{High-efficiency overlay grating fiber-chip couplers for aluminum nitride-on-sapphire waveguide platform,} in \emph{2022 Conference on Lasers and Electro-Optics Pacific Rim (CLEO-PR),}  (2022), pp. 1--2.

\bibitem{smith2022toward}
J.~A. Smith, J.~Monroy-Ruz, P.~Jiang, \emph{et~al.}, \enquote{Toward compact high-efficiency grating couplers for visible wavelength photonics,} {\protect\JournalTitle{Opt. Lett.}} \textbf{47}, 3868--3871 (2022).

\bibitem{ruan2020metal}
Z.~Ruan, J.~Hu, Y.~Xue, \emph{et~al.}, \enquote{Metal based grating coupler on a thin-film lithium niobate waveguide,} {\protect\JournalTitle{Opt. Express}} \textbf{28}, 35615--35621 (2020).

\bibitem{liu2017aluminum}
X.~Liu, C.~Sun, B.~Xiong, \emph{et~al.}, \enquote{Aluminum nitride-on-sapphire platform for integrated high-q microresonators,} {\protect\JournalTitle{Opt. express}} \textbf{25}, 587--594 (2017).

\bibitem{youssefi2021cryogenic}
A.~Youssefi, I.~Shomroni, Y.~J. Joshi, \emph{et~al.}, \enquote{A cryogenic electro-optic interconnect for superconducting devices,} {\protect\JournalTitle{Nat. Electron.}} \textbf{4}, 326--332 (2021).

\bibitem{shen2024photonic}
M.~Shen, J.~Xie, Y.~Xu, \emph{et~al.}, \enquote{Photonic link from single-flux-quantum circuits to room temperature,} {\protect\JournalTitle{Nat. Photon.}} \textbf{18}, 371--378 (2024).

\bibitem{zhao2020design}
Z.~Zhao and S.~Fan, \enquote{Design principles of apodized grating couplers,} {\protect\JournalTitle{J. Lightwave Technol.}} \textbf{38}, 4435--4446 (2020).

\bibitem{lomonte2021efficient}
E.~Lomonte, F.~Lenzini, and W.~H. Pernice, \enquote{Efficient self-imaging grating couplers on a lithium-niobate-on-insulator platform at near-visible and telecom wavelengths,} {\protect\JournalTitle{Optics Express}} \textbf{29}, 20205--20216 (2021).

\bibitem{wang2024heterogeneous}
Y.~Wang, Y.~Guo, Y.~Zhou, \emph{et~al.}, \enquote{Heterogeneous sapphire-supported low-loss photonic platform,} {\protect\JournalTitle{Opt. Express}} \textbf{32}, 20146--20152 (2024).

\bibitem{liu2015smooth}
X.~Liu, C.~Sun, B.~Xiong, \emph{et~al.}, \enquote{Smooth etching of epitaxially grown {AlN} film by {Cl2/BCl3/Ar}-based inductively coupled plasma,} {\protect\JournalTitle{Vacuum}} \textbf{116}, 158--162 (2015).

\bibitem{yeom2005maximum}
J.~Yeom, Y.~Wu, J.~C. Selby, and M.~A. Shannon, \enquote{Maximum achievable aspect ratio in deep reactive ion etching of silicon due to aspect ratio dependent transport and the microloading effect,} {\protect\JournalTitle{J. Vac. Sci. Technol., B: Microelectron. Process. Phenom.}} \textbf{23}, 2319--2329 (2005).

\bibitem{sun2019ultrahigh}
Y.~Sun, W.~Shin, D.~A. Laleyan, \emph{et~al.}, \enquote{Ultrahigh {Q} microring resonators using a single-crystal aluminum-nitride-on-sapphire platform,} {\protect\JournalTitle{Opt. Lett.}} \textbf{44}, 5679--5682 (2019).

\bibitem{martinussen2024thick}
S.~Martinussen, E.~Berenschot, D.~Bonneville, \emph{et~al.}, \enquote{Thick waveguides of low-stress stoichiometric silicon nitride on sapphire (sinos),} {\protect\JournalTitle{Opt. Express}} \textbf{32}, 36835--36847 (2024).

\bibitem{mishra2021mid}
J.~Mishra, T.~P. McKenna, E.~Ng, \emph{et~al.}, \enquote{Mid-infrared nonlinear optics in thin-film lithium niobate on sapphire,} {\protect\JournalTitle{Optica}} \textbf{8}, 921--924 (2021).

\end{thebibliography}

\bibliographyfullrefs{sample}


\end{document}